\documentclass[12pt]{article}
\usepackage {graphicx}
\usepackage [numbers]{natbib}
\def\sun{\hbox{$\odot$}}
\newcommand{\bj}{\mbox{b$_{\rm\scriptscriptstyle J}$}}
\def\aap{A\&A\,  }%% Astronomy and Astrophysics
\def\aj{AJ  }%% The Astronomical Journal
\def\apj{ApJ\,  }%% Astrophysical Journal
\def\apjs{ApJS  }%% Astrophysical Journal, Supplement
\def\bain{BAN  }%% Bulletin of the Astronomical Institutes of the Netherlands 
\def\mnras{MNRAS\,  }%% Monthly Notices of the RAS
\def\rmp{Rev. Mod. Phys.  }% Reviews of Modern Physics
\begin{document}
\begin{center}
\LARGE
\textbf{
New formulae for the  Hubble Constant
in a Euclidean Static Universe
}\\[1cm]
\large
\textbf{Lorenzo Zaninetti}\\[0.5cm]
\normalsize
Dipartimento  di Fisica Generale,
 via P. Giuria 1,
\\ I-10125 Turin,Italy\\
email: zaninetti@ph.unito.it
\\
\end{center}
\medskip
\begin{abstract}
 It is shown that the Hubble constant can be derived
 from the standard luminosity function of galaxies
 as well as from a new luminosity function as
 deduced from the mass-luminosity relationship
 for galaxies.
 An analytical expression for the Hubble constant
 can be found from the
 maximum number of galaxies (in a
 given solid angle and flux) as a function
 of the redshift.
 A second analytical definition of the Hubble constant
 can be found from the redshift averaged 
 over a given solid angle and flux.
 The analysis of two
 luminosity functions for galaxies
 brings to four  the new definitions of the 
 Hubble  constant.
 The equation that regulates the Malmquist bias
 for galaxies is derived and as a consequence it is possible to extract
 a complete sample.
 The application of these  new formulae
 to the data
 of the two-degree Field Galaxy Redshift Survey
 provides a Hubble constant of
 $( 65.26 \pm 8.22 ) \mathrm{\ km\ s}^{-1}\mathrm{\ Mpc}^{-1}$
 for a redshift lower than 0.042.
 All the results  are deduced in a Euclidean universe because 
 the concept of space-time curvature is not necessary
 as well as in a static universe because two 
 mechanisms for the redshift of galaxies alternative
 to the Doppler effect are invoked. 
\end{abstract}
\smallskip
\small
\begin{center}
\textbf{R\'{e}sum\'{e}}
\end{center}
\begin{quote}
\makebox[5mm]{}
Il est montr\'{e}  que la constante de Hubble peut \^{e}tre
d\^{e}riv\'{e}  de la fonction de luminosit\^{e} 
standard pour les galaxies, ainsi que d'une fonction de
luminosit\'{e} nouvelle d\^{e}duite de la relation
masse-luminosit\'{e} pour les galaxies. 
Une expression
analytique de la constante de Hubble peut \^{e}tre
trouv\^{e}e  par rapport au maximum dans le nombre de galaxies
(dans un angle solide donn\'{e} et flux) 
en fonction du
d\^{e}calage vers le rouge . 
Une  deuxi\'{e}me d\^{e}finition analytique
peut \^{e}tre 
trouv\'{e}  par 
la moyenne de  d\^{e}calage vers le rouge  d'un angle solide et le flux. 
Ces deux
d\^{e}finitions sont doubl\^{e}es par l'utilisation d'une
fonction de luminosit\'{e}de nouvelles galaxies.
L'\^{e}quation qui r\^{e}git le biais Malmquist pour les
galaxies est d\^{e}riv\'{e} et avec comme
cons\^{e}quence est possible d'extraire un \^{e}chantillon
complet. L'application de ces nouvelles formules pour les
donn\^{e}es des deux degr\^{e}s Field Galaxy Redshift
Survey fournit une constante de Hubble 
$( 65.26 \pm 8.22 ) \mathrm{\ km\ s}^{-1}\mathrm{\ Mpc}^{-1}$
 pour 
d\^{e}calage vers le rouge 
inf\^{e}rieur \`{a}  0.042.
\end{quote}

%
%\newpage

\section{Introduction}

\begin{figure*}
\begin{center}
\includegraphics[width=10cm]{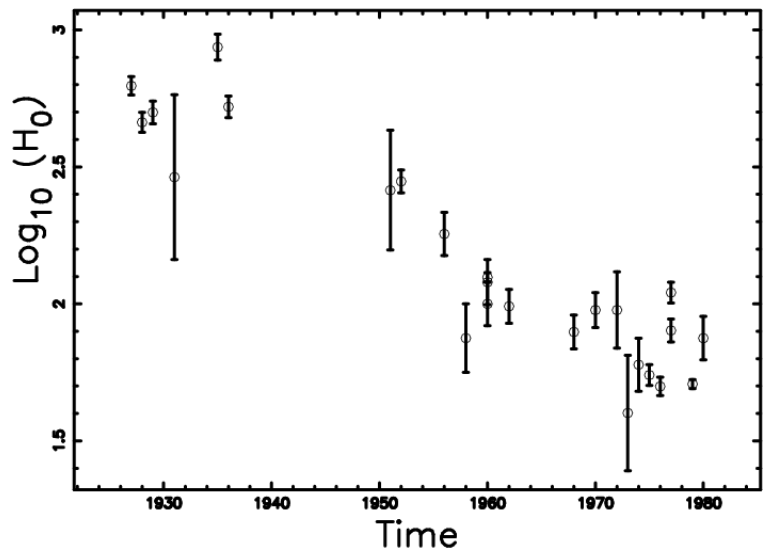}
\end {center}
\caption { Logarithmic values of the Hubble 
constant $H_0$ from 1927 to 1980.
The error bar is evaluated according to the file
http://www.cfa.harvard.edu/~huchra/hubble.plot.dat~.
}
 \label{hystory}%
 \end{figure*}

The Hubble constant, in the following $H_0$, is defined as
\begin{equation}
H_0 = \frac{v}{D} [\mathrm{\ km\ s}^{-1}\mathrm{\ Mpc}^{-1}] \quad
,
\end{equation}
where $v=cz$ is the recession velocity, $D$ is the distance in
$Mpc$, $c$ is the velocity of light and $z$ is the redshift defined
as
\begin{equation}
z = \frac { \lambda_{obs} - \lambda_{em} } { \lambda_{em}} \quad ,
\end{equation}
with $\lambda_{obs}$  and  
$\lambda_{em}$ denoting respectively
 the
wavelengths of the observed and emitted lines
as  determined from the lab source.
The first numerical
values of the Hubble constant
 were :
 $H_{0}=625$ $ \mathrm{\ km\ s}^{-1}\mathrm{\ Mpc}^{-1}$
as deduced by 
\citet{Lemaitre1927}, $H_{0}=460$ $ \mathrm{\ km\
s}^{-1}\mathrm{\ Mpc}^{-1}$ as deduced by
\citet{Robertson1928},
$H_{0}=500 $ $\mathrm{\ km\ s}^{-1}\mathrm{\ Mpc}^{-1}$ as deduced
by  
\citet{Hubble1929}  and $H_{0}=290$ $ \mathrm{\ km\
s}^{-1}\mathrm{\ Mpc}^{-1}$ as deduced by  
\citet{Oort1931}.
Figure~\ref{hystory} reports the decrease of the numerical value
of the Hubble constant from 1927 to 1980.

%begin figure hystory
At the time of writing, two excellent reviews have been written, 
see 
\citet{Tammann2006} $( H_{0}=(63.2\pm1.3~ (random) ~\pm5.3~
(systematic)) $ $ \mathrm{\ km\ s}^{-1}\mathrm{\ Mpc}^{-1} ) $ 
and 
\citet{Jackson2007} 
($ H_0\sim 70\mbox{\,--\,}73$ $ \mathrm{\ km\
 s}^{-1}\mathrm{\ Mpc}^{-1} $).
We now report the methods
that use the global properties of
galaxies as indicators of distance:
\begin{enumerate}
\item Luminosity classes of spiral galaxies;
$H_{0}=(55 \pm 3) $ $\mathrm{\ km\ s}^{-1}\mathrm{\ Mpc}^{-1}$
, see  
\citet{Sandage1999b}
\item 21 cm line widths;
$H_{0}=(59.1 \pm 2.5)$ $ \mathrm{\ km\ s}^{-1}\mathrm{\ Mpc}^{-1}$
,
see 
\citet{Federspiel1999}
\item Brightest cluster galaxies;
$H_{0}=(54.2 \pm 5.4 )$ $ \mathrm{\ km\ s}^{-1}\mathrm{\ Mpc}^{-1}$
,
see 
\citet{Sandage1973}
\item The D$_{n}$-$\sigma$ or fundamental plane method;
$H_{0}=(57 \pm 4)$ $ \mathrm{\ km\ s}^{-1}\mathrm{\ Mpc}^{-1}$
,
see 
\citet{Federspiel1999}

\item Surface brightness fluctuations;
$H_{0}=71.8 $ $\mathrm{\ km\ s}^{-1}\mathrm{\ Mpc}^{-1}$
,
see \citet{Tammann2006}
\item Gravitational lens;
$H_{0}=(72\pm 12) $ $\mathrm{\ km\ s}^{-1}\mathrm{\ Mpc}^{-1}$
,
see \citet{Saha2006}
\item The Sunyaev--Zel'dovich effect;
$H_{0}=(67\pm 18)$ $ \mathrm{\ km\ s}^{-1}\mathrm{\ Mpc}^{-1}$
,
see \citet{Udomprasert2004}
\item Ks-band Tully-Fisher Relation;
$H_{0}=(84 \pm 6 )$ $ \mathrm{\ km\ s}^{-1}\mathrm{\ Mpc}^{-1}$
,
see \citet{Russell2009},
where the Hubble constant was named Hubble parameter.

\end {enumerate}

At the time of writing, the first
important evaluation of the
Hubble constant is
through Cepheids (key programs with HST) and
type Ia Supernovae, see \citet{Sandage2006},
\begin{equation}
H_0 =(62.3 \pm 5 ) \mathrm{\ km\ s}^{-1}\mathrm{\ Mpc}^{-1}
\quad .
\label{h0cefeidi}
\end {equation}
A second  important evaluation
comes from the  three years of  observations with the
Wilkinson Microwave Anisotropy Probe,
see Table 2 in \citet{Spergel2007};
\begin{equation}
H_{0}=(73.2 \pm 3.2)
\mathrm{\ km\ s}^{-1}\mathrm{\ Mpc}^{-1}
\quad .
\label{hzerowmap}
\end{equation}
In the  following,  we will process 
galaxies  having redshifts as given by the catalog 
of galaxies.
The forthcoming analysis is based 
on two key assumptions: (i) the flux of radiation
from galaxies in a given wavelength decreases with 
the square of the distance; (ii) the redshift is assumed
to have a linear relationship with distance in $Mpc$.
These two hypotheses allow   
some new physical mechanisms to be accepted
which produce a linear relationship between redshift 
and distance, for redshifts lower than 1.
In this framework, we can speak of a Euclidean 
universe because the distances are deduced 
from the  Pythagorean theorem  
and a static universe because it is not expanding.
The already listed approaches
leave a series of questions
unanswered or partially answered:
\begin {itemize}
\item
Can the Hubble constant be deduced from
the Schechter luminosity function of galaxies?
\item
Can the Hubble constant be deduced from
a new luminosity of galaxies alternative to 
the Schechter function?
\item Can the equation that regulates the
Malmquist bias be derived in order
to deal with a complete sample in apparent magnitude?
\item
Can the reference magnitude of the sun be deduced from
the luminosity function of galaxies?
\end{itemize}
In order to answer these questions,
Section~\ref{secprelimaries}
contains three introductory paragraphs
on sample moments,
the weighted mean and the determination
of the so-called "exact value" of the Hubble constant.
Section~\ref{useful} reviews the basic system
of magnitudes,
a review of two alternative mechanisms
for the redshift of galaxies, 
two analytical definitions of the
Hubble constant
in terms  of the Schechter luminosity function
of galaxies
and two
other definitions that can be found by adopting a new
luminosity function for galaxies.
Section~\ref{secnumerical} contains a 
numerical evaluation of the four new formulae
for the Hubble constant as deduced from the
data of the
two-degree Field Galaxy Redshift Survey.
Section \ref{secmsun} contains a numerical
evaluation of the reference magnitude of the sun
for a given catalog.

\section{Preliminaries}
\label{secprelimaries}
This Section reviews
the evaluation of the first moment about zero
and of the second moment about the mean of a sample of data,
the evaluation of the mean
and variance when each piece of data of a sample
has differing errors,
the evaluation of the uncertainty
and the evaluation of $H_0$ from a list of published data.

\subsection{Sample moments}

Consider a random sample
${\mathcal X}=x_1, x_2 , \dots , x_n$ and let
$x_{(1)} \geq x_{(2)} \geq \dots \geq x_{(n)}$ denote
their order statistics so that
$x_{(1)}=\max(x_1, x_2, \dots, x_n)$, $x_{(n)}
=\min(x_1, x_2, \dots, x_n)$.
The sample mean, $\bar {x}$ , is
\begin{equation}
\bar {x} = \frac{1}{n} \sum {x_i}
\quad ,
\label{samplemean}
\end{equation}
and the standard deviation of the sample,
$\sigma$ , is according to \citet{press}
\begin{equation}
\sigma = \sqrt{\frac{1}{n-1} \sum (x_i-\bar{x})^2}
\quad .
\label{samplevariance}
\end{equation}

\subsection{The weighted mean }

The probability, $N(x;\mu,\sigma)$,
of a Gaussian (normal) distribution is
\begin{equation}
N(x; \mu,\sigma) =
\frac {1} {\sigma (2 \pi)^{1/2}} \exp {- {\frac {(x-\mu)^2}{2\sigma^2}}}
\quad ,
\label{gaussian}
\end{equation}
where $\mu$ is the mean and $\sigma^2$ the variance.
Consider a random sample ${\mathcal X}=x_1, x_2 , \dots , x_n$
where each value is from a Gaussian distribution
having the same mean but a
different standard deviation $\sigma_i$.
By the
maximum likelihood estimate, in the following MLE
%,
%see 
\cite{Wall2003,Bevington2003}
%,
an estimate of
the weighted mean, $\mu$ ,
is
\begin{equation}
\mu = \frac
{
\sum \frac{xi}{\sigma_i^2}
}
{
\sum \frac{1}{\sigma_i^2}
}
\quad ,
\label{mean}
\end{equation}
and an estimate of the
error of the weighted mean, $\sigma(\mu)$ ,
\begin{equation}
\sigma(\mu) =
\sqrt
{
 \frac
{
1
}
{
\sum \frac{1}{\sigma_i^2}
}
}
\quad ,
\label{error}
\end{equation}
see \cite{Leo1994} for a detailed demonstration.

\subsection{Error evaluation}

When a numerical value of a constant is derived from
a theoretical formula, the
uncertainty  is  found from the error
propagation equation (often called law of errors of Gauss) when
the covariant terms are neglected (see equation (3.14)
in~\cite{Bevington2003}).
In the presence of more than one evaluation of
a constant with different uncertainties,
the weighted mean and
the error of the weighted mean are found
by formulae~(\ref{mean}) and (\ref{error}).
 In the following, in each diagram we will specify
the technique  by which the error bars on the derived
quantities are derived.

\subsection{A first statistical application}

\begin{figure*}
\begin{center}
\includegraphics[width=10cm]{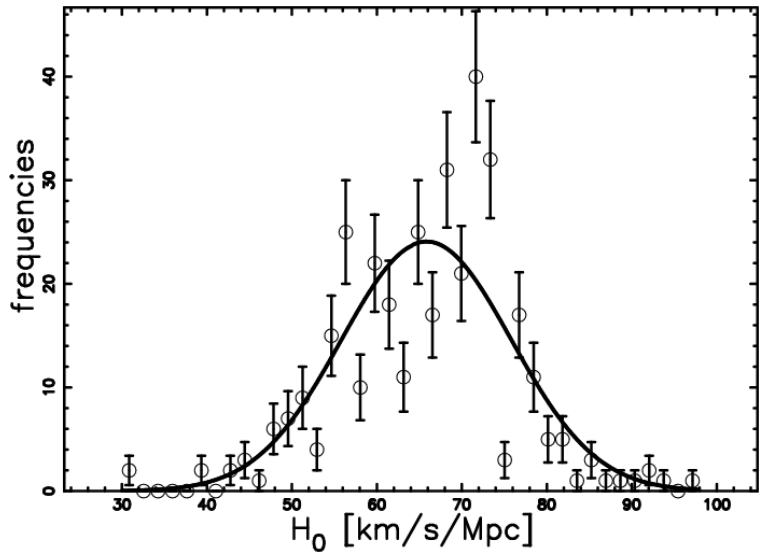}
\end {center}
\caption{ 
	Histogram of frequencies  
  of 355 published values of $H_0$ during the period
1996--2008 with error bars computed as the square root of the
frequencies. The continuous line fit represents a Gaussian
distribution with mean from equation (\ref{mean}) and standard
deviation from equation~(\ref{error}) . }
 \label{gauss}%
 \end{figure*}

The determination of the numerical value of the Hubble
constant is an active field of research and the file
http://www.cfa.harvard.edu/~huchra/hubble.plot.dat
contains a list of 355 published values
during the period 1996--2008.
Figure~\ref{gauss} reports the frequencies
of such values with the superposition of a
Gaussian distribution.
%begin figure gauss

Table~\ref{hubblemany} reports the statistics of this
sample as well the minimum, $H_{0,min}$ and
maximum $H_{0,max}$ .

\begin{table}[ht!]
\caption {
The Hubble constant from
a list of published values
during the period 1996--2008.
}
\label{hubblemany}
\begin{center}
\begin{tabular}{|c|c|c|}
\hline
entity & definition & value   \\
\hline
n & No of samples    & 355 \\
$\bar {x}$ & average & 65.85   $ \mathrm{\ km\
s}^{-1}\mathrm{\ Mpc}^{-1}$ \\
$\sigma $ & standard~deviation & 10 $ \mathrm{\ km\
s}^{-1}\mathrm{\ Mpc}^{-1}$ \\
$H_0,max$ & maximum & 98 $ \mathrm{\ km\
s}^{-1}\mathrm{\ Mpc}^{-1}$ \\
$H_0,min$ & minimum & 30 $ \mathrm{\ km\
s}^{-1}\mathrm{\ Mpc}^{-1}$ \\
$\mu$ & weighted~mean & 66.04  $ \mathrm{\ km\
s}^{-1}\mathrm{\ Mpc}^{-1}$\\
$\sigma(\mu)$ & error~of~the~weighted~mean & 0.25
$ \mathrm{\ km\
s}^{-1}\mathrm{\ Mpc}^{-1}$
 \\
\hline
\end{tabular}
\end{center}
\end{table}
%inizio inserzione
%inizio inserzione
\section{Useful formulae}
\label{useful} This Section reviews three different mechanisms for
the redshifts of galaxies:
the system of
magnitudes, the standard luminosity function
in the following LF
of galaxies and a new LF of galaxies as given by the
mass-luminosity relationship.

\label{formulary}

\subsection{The nature of the redshift}

In the following, we will 
present two theories for the redshift of galaxies 
alternative to the Doppler effect 
which are based on basic axioms of physics.
In these two alternative mechanisms,
the distance, $r$, in a 
Cartesian coordinate system, $x,y,z$,
is given by the usual 
Pythagorean theorem $r=\sqrt {x^2+y^2+ z^2}$.
These two alternative theories do not 
require any expansion of the universe
even though local velocities of
the order of  $\approx~100 \frac{km}{s}$ 
are not excluded.
These random velocities of galaxies  can explain
the bending of radiogalaxies, see \citet{zaninetti2007_b}.

Starting from \citet{Hubble1929}, the suggested correlation between
the expansion velocity and distance in the framework of the Doppler
effect is
\begin {equation}
V= H_0 D = c \, z
\quad ,
\end{equation}
where $H_0$ is the Hubble constant
 $H_0 = 100 h$ $\mathrm{\ km\ s}^{-1}\mathrm{\ Mpc}^{-1}$
, with $h=1$ when $h$ is not specified, $D$ is the distance in
$Mpc$, $c$ is the velocity of light and $z$ the redshift. The
quantity $cz$, a velocity, or $z$, a number, characterizes
the catalog of galaxies. The Doppler effect produces a linear
relationship between distance and redshift. The analysis of 
mechanisms which predict a direct relationship between distance
and redshift started with \citet{Marmet1988} and a current list of
the various mechanisms can be found in \citet{Marmet2009}. Here, 
we select two mechanisms amongst others. The presence of a
hot plasma with low density, such as in the intergalactic medium,
produces a relationship of the type
\begin{equation}
D = \frac{{3.0064 \cdot 10^{24} }}{{ \left( {N_e } \right)_{av}
}}\ln \left( {1 + z} \right)~~{\rm{cm }}{\rm{}} \quad ,
\end{equation}
where  the averaged density of electrons, $\left( {N_e }
\right)_{av} $ , is
\begin{equation}
\left( {N_e } \right)_{av}  = \frac{{H_0}}{{3.076 \cdot 10^5 }}
\approx
 2.42 \cdot 10^{ - 4} \left( \frac{{H_0}}{{74.5}} \right)\,~
{\rm{ cm}}^{ - 3} {\rm{,}}
\end{equation}
see equations (48) and (49) in \citet{Brynjolfsson2004}
or  equation  (27) in  
\citet{Brynjolfsson2009}. 
A second
explanation for the redshift is the Dispersive Extinction Theory
(DET) in which the redshift is caused by the dispersive
extinction of star light by the intergalactic  medium. In this
theory
\begin{equation}
z= (\frac{\pi bc }{4}) \frac{\delta \lambda ^2}
                            {\lambda ^3} D \quad
 ,
\end{equation}
where $\delta \lambda$ is the natural linewidth and $b$ is a
parameter which characterizes the linearity of the extinction,
see formula (17) in \citet{Wang2005}.

\subsection{System of magnitudes}

The absolute magnitude of a galaxy, $M$, is connected
to the apparent magnitude $m$ through the
relationship
\begin{equation}
M = m - 5 Log (\frac {cz}{H_0}) - 25
\quad .
\label{absolute}
\end{equation}
In a Euclidean, non-relativistic
and homogeneous universe,
the flux of radiation,
$ f$, expressed in $ \frac {L_{\sun}}{Mpc^2}$ units,
where $L_{\sun}$ represents the luminosity of the sun,
is
\begin{equation}
f = \frac{L}{4 \pi D_L^2}
\label{flr2}
\quad ,
\end{equation}
where $ D_L$ represents the distance of the galaxy expressed
in $Mpc$
and
\begin{equation}
D_L=\frac{c z}{H_0}
\quad .
\end{equation}

The relationship connecting the absolute magnitude, $M$,
of a
galaxy to its luminosity is
\begin{equation}
\frac {L}{L_{\sun}} =
10^{0.4(M_{\sun} - M)}
\quad ,
\label{mlrelation}
\end {equation}
where $M_{\sun}$ is the reference magnitude
of the sun in the bandpass under consideration.

The flux expressed in $ \frac {L_{\sun}}{Mpc^2}$ units
as a function of the apparent magnitude is
\begin{equation}
f=7.957 \times 10^8 \,{e^{ 0.921\,{\it M_{\sun}}- 0.921\,{\it
m}}}
\quad \frac {L_{\sun}}{Mpc^2} \quad ,
\label{damaf}
\end {equation}
and the inverse relationship is
\begin{equation}
m=M_{\sun}- 1.0857\,\ln \left( 0.1256 \times 10^{-8} f \right)
\quad .
\label{dafam}
\end {equation}
%fine inserzione
%fine inserzione

\subsection{The Schechter function }

The Schechter function, introduced by
\citet{schechter},
provides a useful fit for the
luminosity of galaxies
\begin{equation}
\Phi (L) dL = (\frac {\Phi^*}{L^*}) (\frac {L}{L^*})^{\alpha}
\exp \bigl ( {- \frac {L}{L^*}} \bigr ) dL \quad .
\label{equation_schechter}
\end {equation}
Here, $\alpha$ sets the slope 
for low values of $L$, $L^*$ is the
characteristic luminosity and $\Phi^*$ is the normalization.
The equivalent distribution in absolute magnitude is
\begin{eqnarray}
\Phi (M)dM=&&(0.4 ln 10) \Phi^* 10^{0.4(\alpha +1 ) (M^*-M)}\nonumber\\
&& \times \exp \bigl ({- 10^{0.4(M^*-M)}} \bigr) dM \quad ,
\label{equation_schechter_M}
\end {eqnarray}
where $M^*$ is the characteristic magnitude as derived from the
data.
The joint distribution in {\it z} and {\it f} for galaxies,
see formula~(1.104) in
 \citet{pad}
or formula~(1.117)
in
\citet{Padmanabhan_III_2002},
 is
\begin{equation}
\frac{dN}{d\Omega dz df} =
4 \pi \bigl ( \frac {c}{H_0} \bigr )^5 z^4 \Phi (\frac{z^2}{z_{crit}^2})
\label{nfunctionz}
\quad ,
\end {equation}
where $d\Omega$ , $dz$ and $ df $ represent the differential of
the solid angle, redshift and flux, respectively.
This relationship has been derived assuming
$z \approx \frac{V}{c} \approx \frac {H_0 r}{c}$
and using equation~(\ref{flr2}). 
The critical value of $z$, $z_{crit}$, is
\begin{equation}
 z_{crit}^2 = \frac {H_0^2 L^* } {4 \pi f c^2}
\quad .
\end{equation}

The number of galaxies in $ z$ and $f$ as given by
formula~(\ref{nfunctionz}) has a maximum at $z=z_{pos-max}$ ,
where
\begin{equation}
 z_{pos-max} = z_{crit} \sqrt {\alpha +2 }
\quad ,
\end{equation}
which can be re-expressed as
\begin{equation}
 z_{pos-max} =
\frac
{
\sqrt {2+\alpha}\sqrt {{10}^{ 0.4\,{\it M_{\sun}}- 0.4\,{\it M^*}}}{
\it H_0}
}
{
2\,\sqrt {\pi }\sqrt {f}{\it c}
}
\quad .
\label{zmax_sch}
\end{equation}
From the previous formula, it is possible to derive
a first Hubble constant adopting for the velocity of light
$c=299792.458\mathrm{\frac{km}{s}}$, \citet{CODATA2005},
\begin{eqnarray}
\label{hzero1}
H_0^I=
\frac{N^I}{D^I}
\mathrm{\ km\ s}^{-1}\mathrm{\ Mpc}^{-1} \\
N^I =
{
2.997\times 10^{10}
{\it z_{pos-max}}\,\sqrt {{{\rm e}^{ 0.921{\it
M_{\sun} }- 0.921\,{\it m}}}}
}
 \nonumber \\
D^I =
{
\sqrt { 2+\alpha}\sqrt {{ 10}^{ 0.4\,{\it M_{\sun} }-
 0.4\,{\it M^*}}}
}
\nonumber
\quad .
\end{eqnarray}
The mean redshift of galaxies with a flux $f$,
see formula~(1.105) in~\citet{pad},
or formula~(1.119)
in
\citet{Padmanabhan_III_2002}
 is
\begin{equation}
\langle z \rangle = z_{crit}
\frac {\Gamma (3 +\alpha)} {\Gamma (5/2 +\alpha)}
\quad .
\label{eqnzmedio}
\end{equation}
A second Hubble constant can be derived from the
observed averaged redshift for a given magnitude
\begin{eqnarray}
\label{hzero2}
H_0^{II}=
\frac{N^{II}} {D^{II}}
\mathrm{\ km\ s}^{-1}\mathrm{\ Mpc}^{-1} \\
N^{II} =
1.691\, 10^{10} {\langle z \rangle}_{obs}\times \nonumber \\
\sqrt {\pi }\sqrt {{{\rm e}^{
 0.921\,{\it M_{\sun}}- 0.921\,{\it m }}}}\Gamma
 \left( 5/2+\alpha \right)
\nonumber \\
D^{II} =
{
\Gamma \left( 3+\alpha \right) \sqrt {{10}^{ 0.4\,{\it M_{\sun}}- 0.4\,
{\it M^*}}}
}
\quad ,
\nonumber
\end{eqnarray}
where ${\langle z \rangle}_{obs}$ is the averaged
redshift as evaluated from the considered catalog.

From formula~(\ref{eqnzmedio}), it is also possible
to derive the reference magnitude of the sun
$M_{\sun}$ for the given catalog
\begin{eqnarray}
 M_{\sun} = M^* + \nonumber \\
 1.085\,\ln \left( 1.129\times10^{12}\,{\frac
{{{\it {\langle z \rangle}_{obs}}}^{2}f \left
( \Gamma \left( 2.5+\alpha
 \right) \right) ^{2}}{{{\it H_0}}^{2} \left
( \Gamma \left( 3+
\alpha \right) \right) ^{2}}} \right)
\quad .
\label{msun}
\end{eqnarray}
In this case, $M_{\sun}$ is the unknown and $H_0$ is an
input parameter.

\subsection{The mass-luminosity relationship }

A new LF of galaxies as derived
in \citet{Zaninetti2008} is
\begin{eqnarray}
\Psi (L) dL =&& (\frac{1}{a \Gamma(c_{f}) } ) (\frac {\Psi^*}{L^*})
\left (\frac {L}{L^*} \right )^{\frac{c_{f}-a}{a}} \nonumber\\
&&\times \exp \left ( {-
\left ( \frac {L}{L^*}\right )^{\frac{1}{a}}} \right ) dL \quad
, \label{equation_schechter_mia}
\end {eqnarray}
where $\Psi^*$ is a normalization factor which defines the
overall density of galaxies, a number per cubic $Mpc$,
 $1/a$ is an exponent which connects the mass to the
luminosity
and $c_{f}$ is connected with the dimensionality
of the fragmentation, $c_{f}=2d$, where $d$ represents
the dimensionality of the space being considered: 1,~2,~3.
The distribution
in absolute
magnitude is
\begin{eqnarray}
\Psi (M) dM = && (0.4 ln 10 \frac {1}{a \Gamma(c_{f})}) \Psi^*
10^{0.4(\frac{c_{f}}{a}) (M^*-M)} \nonumber \\
&& \times \exp \bigl ({-
10^{0.4(M^*-M)(\frac{1}{a})}} \bigr) dM \quad.
\label{equation_mia}
\end {eqnarray}
This function contains the parameters $M^*$, {\it a},
$c_{f}$
and $\Psi^*$ which are  derived from the operation of fitting
the experimental data.
The joint distribution in $z$ and $f$,
in the presence of the ${\mathcal M}-L$
luminosity (equation~(\ref{equation_schechter_mia})) is
\begin{equation}
\frac{dN}{d\Omega dz df} =
4 \pi \bigl ( \frac {c}{H_0} \bigr )^5 z^4 \Psi (\frac{z^2}{z_{crit}^2})
\label{nfunctionz_mia}
\quad .
\end {equation}
The number of galaxies, $N_{{\mathcal M}-L}(z,f_{min},f_{max})$
comprised between
 $f_{min}$ and $f_{max}$,
can be computed through the following integral
\begin{equation}
N_{{\mathcal M}-L} (z) = \int_{f_{min}} ^{f_{max}}
4 \pi \bigl ( \frac {c}{H_0} \bigr )^5 z^4 \Psi (\frac{z^2}{z_{crit}^2})
df
\quad ,
\label{integrale_mia}
\end {equation}
and also in this case
a numerical integration must be performed.

The number of galaxies as given
by formula~(\ref{nfunctionz_mia}) has a maximum at
$z_{pos-max}$ where
\begin{equation}
 z_{pos-max} = z_{crit}
\left( {\it c_{f}}+a \right) ^{a/2}
\quad ,
\end{equation}
which can be re-expressed as
\begin{equation}
 z_{pos-max} =
\frac
{
\left( a+{\it c_{f}} \right) ^{1/2\,a}\sqrt {{10}^{ 0.4\,{\it M_{\sun}}-
0.4\,{\it M^*}}}{\it H_0}
}
{
2\,\sqrt {\pi }\sqrt {f}{\it c}
}
\quad .
\label{zmax_mia}
\end{equation}
A third Hubble constant as deduced from
the maximum in the number of galaxies as a function of z
is
\begin{eqnarray}
\label{hzero3}
H_0^{III}=
\frac {N^{III} } {D^{III}}
\mathrm{\ km\ s}^{-1}\mathrm{\ Mpc}^{-1}
 \\
N^{III} =
2.997 \times 10^{10} {\it z_{pos-max}}\,\sqrt {{{\rm e}^{ 0.921\,{\it
M_{\sun} }- 0.921 {\it m }}}}
 \\
D^{III}=
{
\left( {\it c_{f}}+a \right) ^{ 0.5\,a}\sqrt {{ 10.0}^{
 0.4\,{\it M_{\sun}}- 0.4\,{\it M^*}}}
}
\nonumber
\quad .
\end{eqnarray}
The mean redshift
connected with the ${\mathcal M}-L$ LF
 is
\begin{equation}
\langle z \rangle = z_{crit}
\frac {2\;\;{4}^{-{\frac {2\,a+{\it c_{f}}}{a}}}\Gamma
\left( 2\,a+{\it c_{f}}
 \right) {2}^{{\frac {2\,{\it c_{f}}+3\,a}{a}}} }{ \Gamma \left( {
\it c_{f}}+3/2\,a \right) }
\quad
\label{zmediomia}
\end{equation}
and the fourth Hubble constant is
\begin{eqnarray}
\label{hzero4}
H_0^{IV}=
\frac{N^{IV}} {D^{IV}}
\mathrm{\ km\ s}^{-1}\mathrm{\ Mpc}^{-1}
 \\
N^{IV}=
8.457\, 10^{9} {\it {\langle z \rangle}_{obs} } \times
\nonumber \\
\sqrt {\pi }\sqrt {{{\rm e}^{
 0.921\,{\it M_{\sun} }- 0.921,{\it m }}}}\Gamma
 \left( {\it c_f }+3/2\,a \right)
\nonumber \\
D^{IV}=
{4}^{-{\frac {2\,a+{\it c_{f}}}{a}}}\sqrt {{10}^{ 0.4\,{\it M_{\sun}}- 0.4
\,{\it M^*}}} \times
\nonumber \\
\Gamma \left( 2\,a+{\it c_{f}} \right) {2}^{{\frac {2\,{
\it c_{f}}+3\,a}{a}}}
\nonumber
\quad .
\end{eqnarray}

\section{Numerical value of the Hubble constant}

\label{secnumerical}
\begin{figure*}
\begin{center}
\includegraphics[width=10cm]{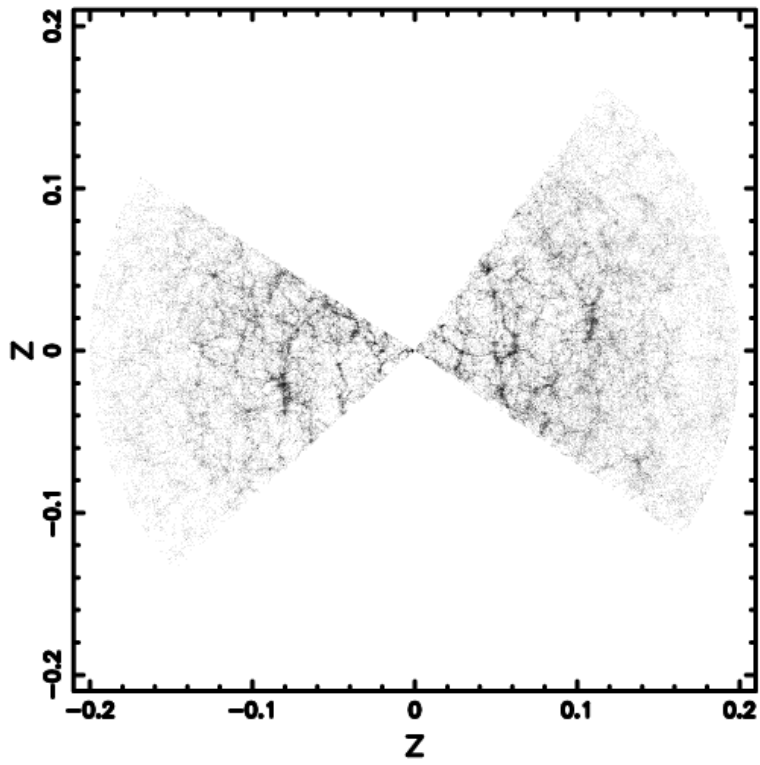}
\end {center}
\caption{Cone-diagram  of all the  galaxies  
in the 2dFGRS.
This plot contains  203,249  galaxies.  
}
          \label{2df_all}%
    \end{figure*}

\begin{figure}
\begin{center}
\includegraphics[width=10cm]{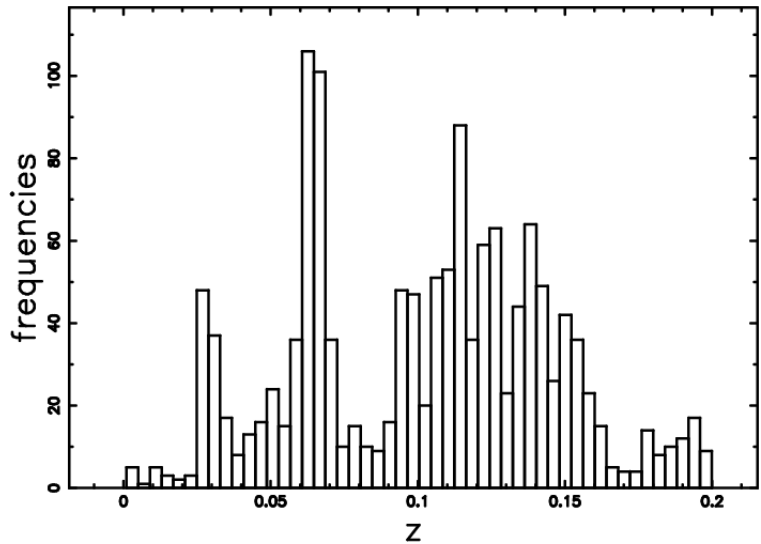} 
\end {center}
\caption {
Histogram (step-diagram) of the 
number of galaxies as a function of the 
redshift in the  slice to the right of Figure
\ref{2df_all}, the number of bins is 50.
The  circular sector has a central angle of
$1^\circ$. 
}
\label{isto2}
    \end{figure}

The formulae previously derived are now tested
on the catalog from the
two-degree Field Galaxy Redshift Survey,
in the following 2dFGRS,
available at the web site: http://msowww.anu.edu.au/2dFGRS/.
In particular we added together the file parent.ngp.txt which
contains 145,652 entries for NGP strip sources and
the file parent.sgp.txt which
contains 204,490 entries for SGP strip sources.
Once the heliocentric redshift was selected,
we processed 219,107 galaxies with
$0.001 \leq z \leq 0.3$
and  two strips of  the 2dFGRS are 
shown in Figure~\ref{2df_all}.
%begin figure 2df_all
%end   figure 2df_all
From the  previous Figure is  clear the nonhomogeneous 
structure of the universe and this concept
can be clarified by counting the number of galaxies
in one of the two slices 
as a function of the redshift when a sector with
a central angle of $1^\circ$ is considered,
see Figure~\ref{isto2}.

%begin figure isto2
\begin{figure}
\begin{center}
\includegraphics[width=10cm]{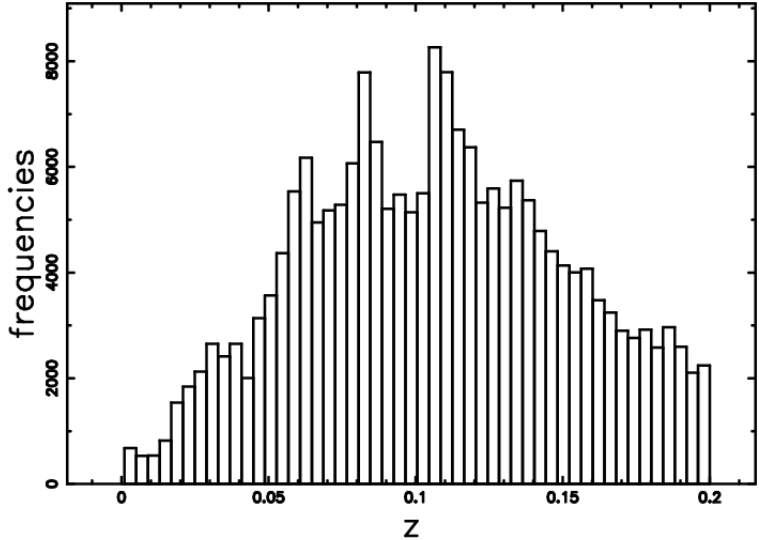} 
\end {center}
\caption {
Histogram (step-diagram) of the 
number of galaxies as a function of the 
redshift when the two slices of Figure
\ref{2df_all} are added together, 
the number of bins is 50.
}
\label{isto1}
    \end{figure}

% end figure isto2
Conversely, when the two slices are considered together
the behavior of the number of galaxies 
as a function of the redshift is more continuous,
see Figure~\ref{isto1}. 

%begin figure isto1
\begin{figure*}
\begin{center}
\includegraphics[width=10cm]{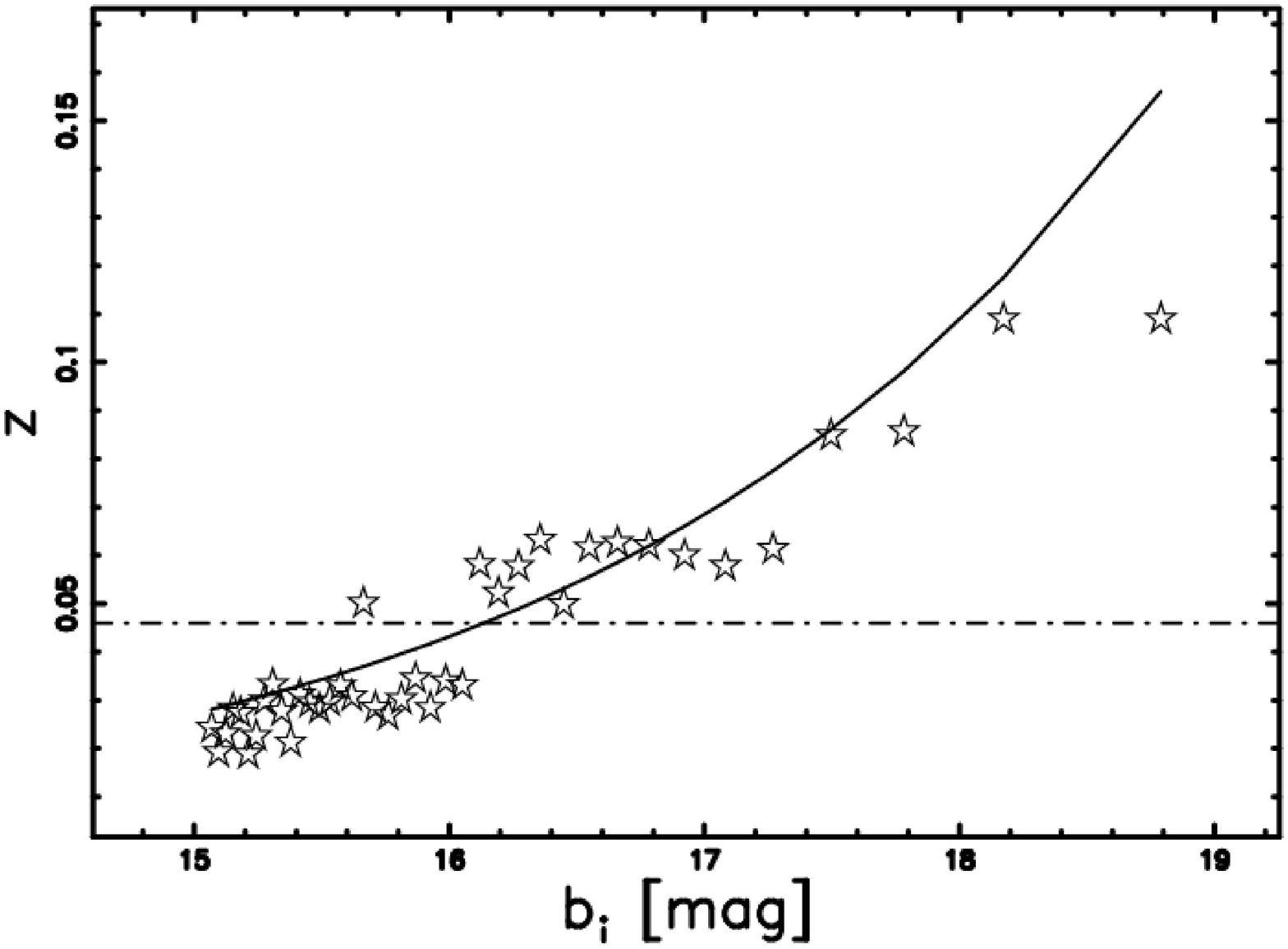}
\end {center}
\caption{
Value of  
$\widehat z_{pos-max}$ 
 at which the number of
 galaxies in the 2dFGRS 
is maximum as a function of 
the apparent magnitude $\bj$ 
(stars) and 
theoretical curve of the maximum for the 
Schechter function  
as represented by  formula~(\ref{zmax_sch}) 
(full line).
In this plot, $\mathcal{M_{\sun}}$ = 5.33  
and $H_0 = 65.26 \mathrm{\ km\ s}^{-1}\mathrm{\ Mpc}^{-1}$.
The horizontal dotted line represents the boundary
between complete and incomplete samples.
}
          \label{zeta_max_flux}%
    \end{figure*}

% end figure isto1
In this  quasi-homogeneous universe,  
some statistical
properties such as 
the theoretical 
position of the 
maximum in the number of galaxies
agree with the observations and 
Figure~\ref{zeta_max_flux} reports 
the observed  maximum  in the 2dFGRS
as well as
the theoretical curve as a function of the magnitude.
%begin figure zeta_max_flux

\begin{figure*}
\begin{center}
\includegraphics[width=10cm]{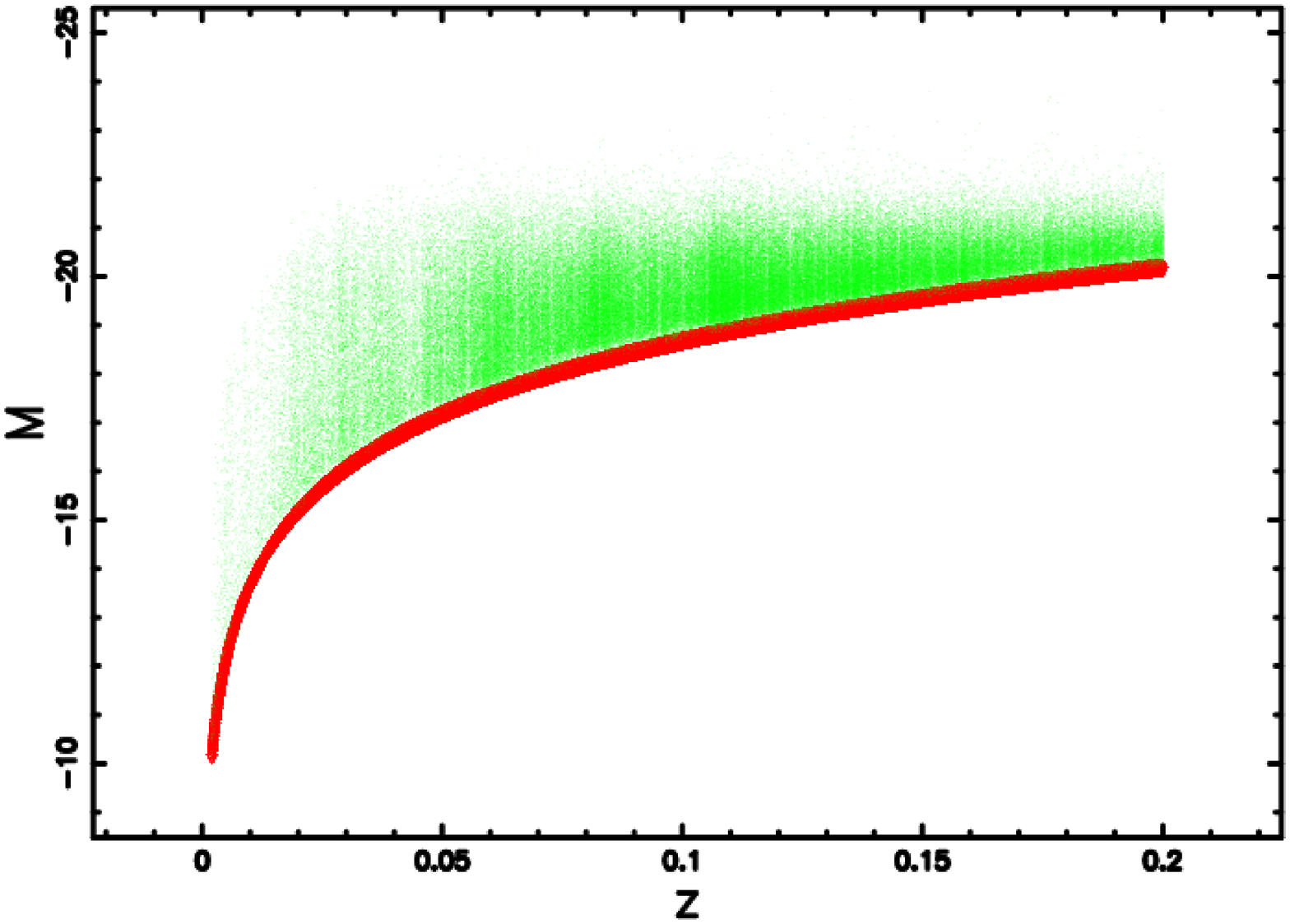}
\end {center}
\caption{
The absolute magnitude $M$ of
202,923 galaxies belonging to the 2dFGRS
when $\mathcal{M_{\sun}}$ = 5.33 and
$H_{0}=66.04 \mathrm{\ km\ s}^{-1}\mathrm{\ Mpc}^{-1}$
(green points).
The upper theoretical curve as represented by
equation~(\ref{absolutel}) is reported as the
red thick line when $m_L$=19.61.
}
 \label{bias}%
 \end{figure*}

Before reducing the data, we should
discuss the Malmquist bias,
see \citet{Malmquist_1920,Malmquist_1922},
which was originally applied
to the stars and was 
then applied to the galaxies by \citet{Behr1951}. We
therefore introduce
 the concept of
limiting apparent magnitude and the corresponding
 completeness in
absolute magnitude of the considered catalog as a function of the
redshift. The observable absolute magnitude as a function of the
limiting apparent magnitude, $m_L$, is
\begin{equation}
M_L =
m_{{L}}-5\,{\it \log_{10}} \left( {\frac {{\it c}\,z}{H_{{0}}}}
 \right) -25
\quad .
\label{absolutel}
\end{equation}
The previous formula predicts, from a theoretical
point of view, an upper limit on the absolute
maximum magnitude that can be observed in a
catalog of galaxies characterized by a given limiting
magnitude and Figure~\ref{bias} reports such a curve
and the galaxies of the
2dFGRS.
%begin figure bias

The interval covered by the 
LF of galaxies,
$\Delta M $,
is defined by 
\begin{equation}
\Delta M = M_{max} - M_{min}
\quad ,
\end{equation}
where $M_{max}$ and $M_{min}$ are the
maximum and minimum
absolute
magnitude of the LF for the considered catalog.
The real observable interval in absolute magnitude,
$\Delta M_L $,
 is
\begin{equation}
\Delta M_L = M_{L} - M_{min}
\quad .
\end{equation}
We can therefore introduce the range
of observable absolute maximum magnitude
expressed in percent,
$ \epsilon_s(z) $,
as
\begin{equation}
\epsilon_s(z) = \frac { \Delta M_L } {\Delta M } \times 100
\, \% 
\quad .
\label{range}
\end{equation}
This is a number that represents 
the completeness
of the sample
and, given the fact that the limiting magnitude of the 2dFGRS is
$m_L$=19.61, it is possible to conclude that the 2dFGRS is complete
for $z\leq0.0442$~. 
This efficiency expressed as a 
percentage 
can be considered a version  of the Malmquist bias. 
In our case, we have chosen to process
the galaxies of the 2dFGRS with $z\leq0.0442$ of which there are
 22,071: in other words our sample is complete.
 Another quantity that should be fixed in order
to continue is
the absolute magnitude of
the sun in the $\bj$ filter,
$\mathcal{M_{\sun}}$ = 5.33, see
\citet{Colless2001,Einasto_2009,Eke_2004}.

\begin{figure*}
\begin{center}
\includegraphics[width=10cm]{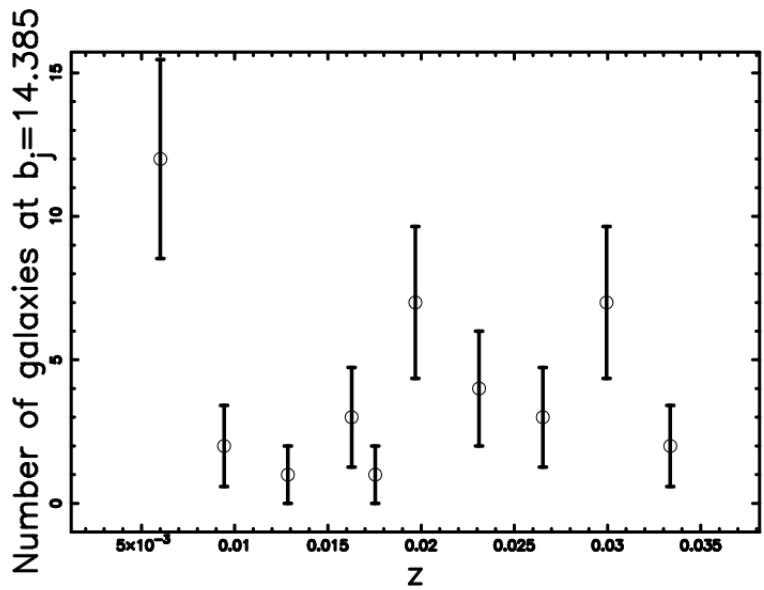}
\end {center}
\caption{
The galaxies of the 2dFGRS, with
$ \bj \approx 14.385 $ or $f$
 $\approx
189983 \frac {L_{\sun}}{Mpc^2}$,
are isolated
in order to represent a chosen value of $m$ or $f$
and then organized according to frequency versus
heliocentric redshift.
The error bars are computed as the square root of the
frequencies.
The maximum in the frequency of observed galaxies is
at $z=0.006$ when $\mathcal{M_{\sun}}$ = 5.33 .
}
 \label{istogram}%
 \end{figure*}

\begin{figure*}
\begin{center}
\includegraphics[width=10cm]{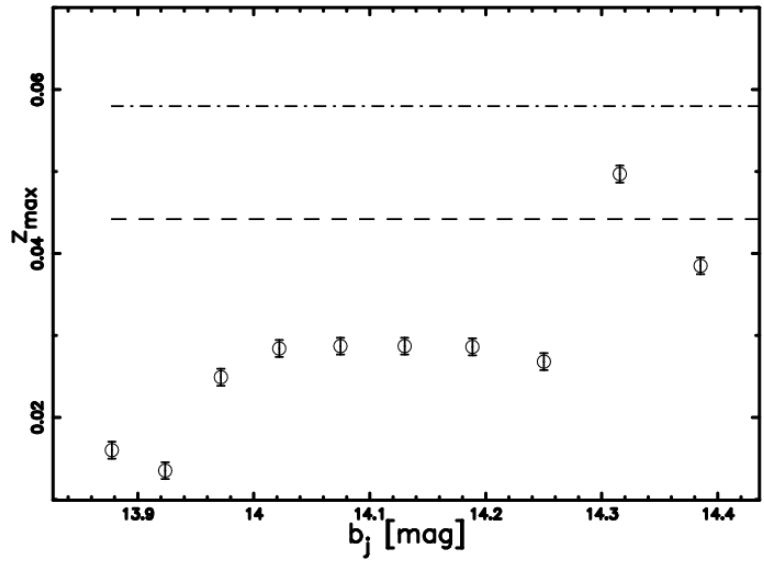}
\end {center}
\caption{
Plot of $z_{max}$ as a function of the chosen magnitude
(empty stars).
The error bar in $z$ is computed as the width of the bin.
The dashed line represents the lower limit
of the complete sample,
$ \epsilon_s(z) =100\%$, and the
dash-dot-dash line corresponds to $ \epsilon_s(z) =90\%$.
}
 \label{sample}%
 \end{figure*}

We now outline the algorithm that allows
to deduce
$z_{pos-max}$ and
${ \langle z \rangle}_{obs} $ from a catalog of galaxies.
\begin{enumerate}
\item
We fix a given flux or magnitude, for
example $\bj$, and a relative narrow window.
\item
We organize the selected galaxies according to frequency
versus redshift, see a typical histogram in
Figure~\ref{istogram}.
\item
Once the histogram is made, we compute
the astronomical $z=z_{pos-max}$, which is inserted in formulae
(\ref{hzero1}) and (\ref{hzero3}) in order to deduce the
Hubble constant.
\item The selected sample of galaxies with a given magnitude
allows an easy determination of
${\langle z \rangle}_{obs} $.
\item Particular attention should be paid
to the completeness
of the sample and Figure~\ref{sample} reports the
maximum value in redshift $z_{max}$ for each run
in magnitude/flux.
\end{enumerate}

\begin{figure*}
\begin{center}
\includegraphics[width=10cm]{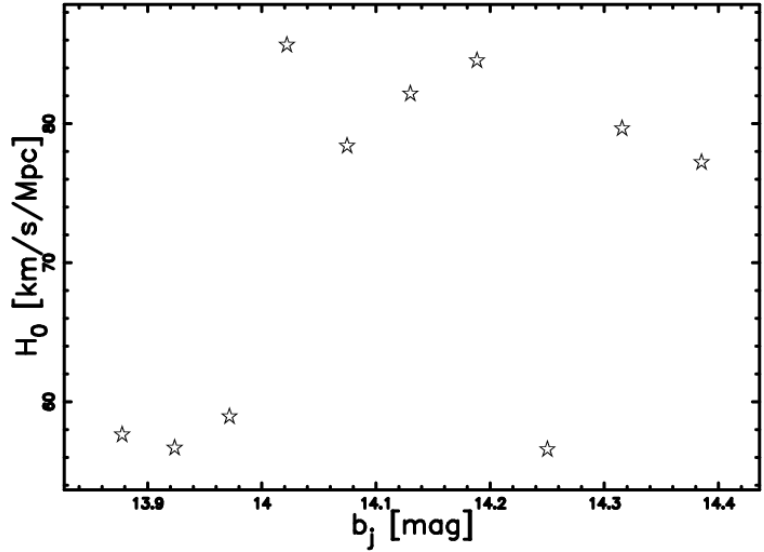}
\end {center}
\caption
{
The Hubble constant as deduced by
the second method, see equation~(\ref{hzero2}),
as a function of the selected magnitude (empty stars).
}
 \label{hzerofig}%
 \end{figure*}

Table~\ref{hubblevalue} reports the four values
of the Hubble constant deduced here
and Figure~\ref{hzerofig} displays the data
corresponding to the constant deduced
from equation~(\ref{hzero2}).

%begin figure istogram

%begin figure sample

%begin figure hzerofig

\begin{table}[ht!]
\caption {
Numerical values of the Hubble constant
as deduced from 10 different apparent
magnitudes.
}
\label{hubblevalue}
\begin{center}
\begin{tabular}{|c|c|c|c|}
\hline
~ & LF & matching~$z$ & [$
\mathrm{\ km\ s}^{-1}\mathrm{\ Mpc}^{-1}]
 $
\\
\hline 1 & Schechter & $ z_{pos-max}$ & ( 58.35 $\pm$ 30 )
 \\
2 & Schechter & $ {\langle z \rangle}_{obs}$ & ( 71.73$\pm$ 12)
 \\
3 & ${\mathcal M}-L$ & $ z_{pos-max}$ & ( 60.72 $\pm$ 32)
 \\
4 & ${\mathcal M}-L$ & $ {\langle z \rangle}_{obs}$ & ( 71.20
$\pm$ 12 )
 \\
5 & weighted~mean & ~ & ( 65.26 $\pm$ 8.22 )
 \\
6 & sample~mean & ~ & (62.88 $\pm$ 6.0 )
\\
\hline
\end{tabular}
\end{center}
\end{table}
From a practical point of view,
$\epsilon$,
the percentage 
reliability of our
results can also be introduced,
\begin{equation}
\epsilon =(1- \frac{\vert( Q_{obs}- Q_{num}) \vert}
{Q_{obs}}) \cdot 100
\, \% 
\,,
\label{efficiency}
\end{equation}
where $Q_{obs}$ is the quantity given
by the astronomical observations 
and
$Q_{num}$ is the analogous
quantity calculated by us.
The value of $H_0$ as found by us with the weighted mean is,
see fifth row in Table~\ref{hubblevalue},
$H_{0}=65.26 \mathrm{\ km\ s}^{-1}\mathrm{\ Mpc}^{-1}$
and the observed value,
see the weighted mean in Table~\ref{hubblemany},
$H_{0}=66.04 \mathrm{\ km\ s}^{-1}\mathrm{\ Mpc}^{-1}$ .

\section{The absolute magnitude of the sun}
\label{secmsun}
\begin{figure*}
\begin{center}
\includegraphics[width=10cm]{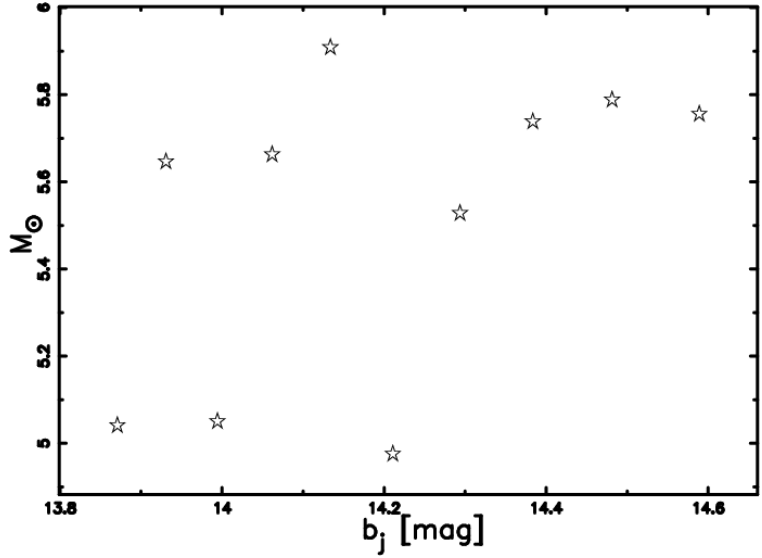}
\end {center}
\caption
{
The absolute reference magnitude of
the sun, equation~(\ref{msun}),
as a function of the selected magnitude (empty stars).
}
 \label{sun}%
 \end{figure*}

The reference absolute magnitude of the sun
(the unknown variable)
can be derived from formula~(\ref{msun})
but in this case the value of $H_0$
(known variable) should be specified.
Perhaps the best choice is the weighted mean
reported in Table~(\ref{hubblemany}),
$H_0=66.04 \mathrm{\ km\ s}^{-1}\mathrm{\ Mpc}^{-1}$.
Adopting this value of $H_0$, the absolute
reference magnitude of
the sun can be plotted in Figure~\ref{sun}
and the averaged value is
\begin{equation}
\overline{M_{\sun}} = (5.50\pm 0.35) mag \quad . \label{msunvalue}
\end{equation}
%begin figure sun
The efficiency in deriving the absolute reference magnitude
of the sun is
\begin{equation}
\epsilon = 96.63 ~\% \quad .
\end{equation}

\section{Conclusions}

A careful study of the standard LF of galaxies
allows the determination
of the position of the maximum
in the theoretical number of galaxies versus redshift
and the theoretical averaged
redshift.
From the two previous analytical results, it is possible
to extract two new formulae for
the Hubble constant, equations~(\ref{hzero1}) and
(\ref{hzero2}).
The same procedure can be applied by analogy
to a new LF as given by the mass-luminosity
relationship, see
equations~(\ref{hzero3}) and
(\ref{hzero4}).
The weighted mean
of the four  values of $H_0$ as deduced from
Table~\ref{hubblevalue} gives

\begin{equation}
H_0 =( 65.26 \pm 8.22  ) \mathrm{\ km\ s}^{-1}\mathrm{\ Mpc}^{-1}
\quad when~ z \leq 0.042 \nonumber \quad .
\end {equation}
This value lies between the
value deduced from the Cepheids,
see~\citet{Sandage2006} and formula~(\ref{h0cefeidi})
and the value deduced from WMAP,

see~\citet{Spergel2007} and formula~(\ref{hzerowmap}).

The developed framework also enables  
the deduction of the reference magnitude of the sun,
see formula~(\ref{msun}) and
the application to the 2dFGRS
gives
\begin{equation}
 M_{\sun} = (5.5 \pm 0.35)
\quad .
\end{equation}

Assuming that the exact value is $ M_{\sun}$ = 5.33, 
the efficiency
in deriving the reference magnitude of the sun is
$\epsilon=96.63
~\% $ when $H_0=66.04 \mathrm{\ km\ s}^{-1}\mathrm{\ Mpc}^{-1}$.
%inizio inserire
%inizio inserire
We briefly review the basic cosmological 
assumptions adopted here 
to derive the Hubble constant:
\begin{itemize}
\item The mechanism that produces the redshift, 
 here extracted
 from the catalog of galaxies,
 is not specified but we remember
 that the plasma redshift and DET 
 (Dispersive Estinction Theory) 
do not produce
 a geocentric
 model for the universe as given by the Doppler shift, see
 \citet{Wang2007}.
 
\item The number of galaxies as a function 
      of redshift as
      well as the averaged redshift are evaluated 
      in a Euclidean space or, in other words,
      the effects of space-curvature are ignored. 
 \item
The spatial inhomogeneities present in the catalog of galaxies
are partially neutralized by  the operation of adding together
the data of the south and north galactic pole of the 2dFGRS.
The transition from a nonhomogeneous  
to a quasi-homogeneous universe
is clear when Figure~\ref{isto1} and Figure~\ref{isto2}
are carefully analyzed. 
\item 
The  initial  assumptions of: 
(i) natural  flux decreasing
as given by equation~(\ref{flr2}) ; 
(ii) linear relationship
between redshift and distance 
which are present 
 in the joint distribution in {\it z} and {\it f} 
for the number of galaxies  are  justified 
by the acceptable results obtained for 
the theoretical maximum in the number of galaxies, 
see Figure~\ref{zeta_max_flux}.
This fact allow us to speak of a 
Euclidean universe up to $z \leq 0.042$.
\item
The presence of the Malmquist bias 
does not allow to 
extrapolate the concept of a Euclidean, static universe
for distances greater than $z\,>\,0.042$
when the 2dFGRS  catalog is considered.
%siamoqui
\end{itemize}
%fine inserire
%fine inserire

\section*{Acknowledgments}
 I would like to thank
 the Smithsonian Astrophysical Observatory and
 John Huchra for the public file
 http://www.cfa.harvard.edu/~huchra/hubble.plot.dat
 which contains the published values of the Hubble constant.

%\nocite{*}
%\bibliography{biblio}

\end{document}